# CXL over Ethernet: A Novel FPGA-based Memory Disaggregation Design in Data Centers

Chenjiu Wang*
wangchenjiu20s@ict.ac.cn

Ke He*
pixiesunky@gmail.com

Ruiqi Fan
fanruiqi20g@ict.ac.cn

Xiaonan Wang
wangxiaonan@iicit.net

Yang Kong
kongyang84@gmail.com

Wei Wang
herry.wangwei@gmail.com

Qinfen Hao
haoqinfen@ict.ac.cn

*Abstract*—Memory resources in data centers generally suffer from low utilization and lack of dynamics. Memory disaggregation solves these problems by decoupling CPU and memory, which currently includes approaches based on RDMA or interconnection protocols such as *Compute Express Link (CXL)*. However, the RDMA-based approach involves code refactoring and higher latency. The CXL-based approach supports native memory semantics and overcomes the shortcomings of RDMA, but is limited within rack level. In addition, memory pooling and sharing based on CXL products are currently in the process of early exploration and still take time to be available in the future. In this paper, we propose the *CXL over Ethernet* approach that the host processor can access the remote memory with memory semantics through Ethernet. Our approach can support native memory load/store access and extends the physical range to cross server and rack levels by taking advantage of CXL and RDMA technologies. We prototype our approach with one server and two FPGA boards with 100 Gbps network and measure the memory access latency. Furthermore, we optimize the memory access path by using data cache and congestion control algorithm in the critical path to further lower access latency. The evaluation results show that the average latency for the server to access remote memory is 1.97 μs, which is about 37% lower than the baseline latency in the industry. The latency can be further reduced to 415 ns with cache block and hit access on FPGA.

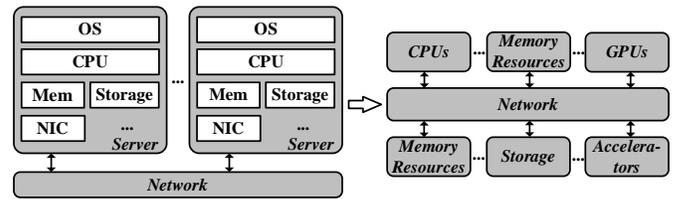

Fig. 1. The architecture of the disaggregation system.

## I. Introduction

Data centers and cloud services play an important role in applications such as big data analytics, deep learning, and graph computing, which leads to a huge demand for memory [1]. However, in many cases, memory resources in data centers generally suffer from low memory utilization and lack of dynamics.

Low memory utilization mainly includes low usage after memory allocation and memory stranding. The former refers to the scenarios in which the application requests a large amount of memory, but the memory is not fully used in practice, and thus it cannot be allocated and used by other applications. The memory stranding problem refers to when all the CPU cores of the server are rented to virtual machines (VMs), there may still be a portion of memory that has not been allocated to any VM and will not be used for a long time. Studies have shown that the percentage of memory that is actually actively used in data centers is about 50%, and about 30% of the overall memory is allocated but not used [2]. An average of 6% of memory in the data center is stranded when 75% of CPU cores are scheduled for VMs, and when the number of VMs or containers increases, the average stranded memory also increases and will reach a maximum of 25% [3].

The lack of memory dynamics refers to the higher requirements for data center memory in cloud-native serverless scenarios with fast scaling, second-level elasticity, high density, and high capacity [4]. However, limited by the fixed CPU and memory ratios of the current server architecture, it is difficult to dynamically scale the memory capacity of a single server to the required size, and the server cannot directly use the memory of other servers. As a result, fast elastic scaling and high-density memory instances cannot be achieved.

To solve these problems, the disaggregation architecture has been proposed, which decouples various physical resources in a server to form resource pools. Fig. 1 shows the architecture of the disaggregation system. For example, GPUs for heterogeneous computing, storage resources, and memory resources can be physically separated, which can be allocated and expanded on demand through high-speed network interconnection and control interface, thus further improving hardware resource utilization and reducing data center TCO [5]. To solve the problem of memory resource utilization and dynamics, we focus on the memory disaggregation system which organizes CPU and memory as compute nodes (CNs) and memory nodes (MNs) respectively. There are two main approaches to implementing a memory disaggregation system. One is to build distributed memory systems that provide global memory management through distributed applications running on individual servers [6-10]. They usually provide the fine

---

granularity of memory sharing. The other use memory pools composed of raw memory devices [11-13]. Memory pools can provide extended memory for individual CNs, which are independent of each other and usually do not share data. This paper mainly focuses on the latter approach.

The memory disaggregation systems can be classified by physical transmission approach from CN to MN as RDMA-based [1, 6, 7, 10, 13-18] or interconnection protocols such as CXL that support cache coherence based [19-23]. RDMA-based systems inherit the disadvantages of RDMA [24], i.e., RDMA memory accesses do not natively support memory semantics, so applications need to explicitly call the relevant APIs [25], and manage MR (Memory region) and QP (Queue pair) [26] through the runtime. In addition, the RDMA approach needs the RNIC (RDMA Network Interface Card) driver to perform address translation between MR and physical memory, as well as DMA copies between memory and NIC, which increase latency and interrupt overhead. In summary, RDMA memory access adds intrusive code modification and OS system call overhead, which increases latency compared to using native memory semantics. According to the data exchange granularity between CN and MN, RDMA-based designs can be further classified into page-based [1, 14-16] or object-based [6, 7, 10, 13, 17, 18], both of which have additional overheads. For example, page-based systems involve additional page faults and read/write data amplification. Object-based systems require custom APIs, which sacrifices transparency.

Memory disaggregation based on interconnection protocols such as CXL that support cache coherence natively supports memory semantics and therefore overcomes the drawbacks of RDMA. They are usually implemented with CC-FPGAs (cache coherence FPGAs), which can be connected to the CPU through the interface that implements cache coherence protocol and supports memory sharing between CPU and FPGA. Various platforms have been proposed, such as OpenCAPI [27] for IBM, CCIX [28] for AMD/ARM, and CXL [29] for Intel. Using these protocols, the CPU can access FPGA-attached memory using load/store semantics, and data communication between them is at cache line granularity. The access latency is similar to NUMA nodes, and applications need no code changes. However, this approach is limited within rack level even for the latest CXL 3.0 specification, while RDMA can access memory between servers across racks. Some studies [19, 20] have tried to extend the distance of interconnection protocols, but they still require additional RDMA operations.

In summary, existing memory disaggregation systems cannot achieve all aspects of low latency, fine-grained memory management, scalability, and application transparency. Therefore, we propose the *CXL over Ethernet* approach. We use the CXL-enabled CC-FPGA and take full advantage of the memory semantics supported by CXL to access remote memory as if it were a NUMA node, thus avoiding the increased latency and system overhead associated with RDMA. Our approach is transparent to the application, does not require additional drivers, and implements cache line granularity for data exchange. Considering scalability, We use Ethernet to transmit CXL remote memory access requests, which overcomes the drawback that native CXL is limited to the scope of the rack. CXL memory requests from the host can be sent directly to FPGA, and FPGA interacts with the remote memory through its network interface. We encapsulate Ethernet frames on FPGA, which is transparent to the kernel and does not go through the traditional network stack, thus avoiding DMA copies and address translation using the RDMA approach.

In addition, there are bandwidth and latency issues to consider. First, the current Ethernet bandwidth is gradually moving toward 800 Gbps or even 1.6 Tbps [30], which is sufficient to meet transmission bandwidth requirements. Second, the CXL latency on the host side is already less than RDMA. We further add a cache on the FPGA of CN side to cache the remote memory data, which reduces the frequency of using Ethernet and the latency of the network path. Few systems that use CC-FPGAs consider network congestion control, and we design a switch-independent congestion control algorithm to reduce the latency increase due to network congestion.

In this paper, we propose *CXL over Ethernet* approach to build a memory disaggregation system and implement the prototype. Our contributions are as follows.

- We utilize the emerging CXL as the cache coherence protocol and work with customized Ethernet frame format to access remote memory with native memory semantics. To our best knowledge, it is the first time to implement a memory disaggregation system on multiple FPGA boards.

- Compared to prior work, our approach firstly introduces local FPGA cache and congestion control algorithm which can lower remote memory access latency and increase reliability in data center scenarios.

- Our system achieves an average remote memory access latency of 1.97 μs under round-trip latency tests with 64 bytes, 37% latency saving compared to the RDMA-based approaches. Furthermore, we can achieve the remote memory latency of 415 ns with cache logic when all memory accesses are cache hit.

- Our system has demonstrated that it can combine the benefits of both RDMA and native CXL technologies so that the software has no code refactoring to dynamically expand memory without physical limitations at the same time, which is viewed as one of the most essential factors in data center scenarios.

## II. RELATED WORK

### A. Memory Disaggregation System

Current memory disaggregation systems mainly use the traditional RDMA approach or use cache coherence protocols such as CXL.

**Memory Disaggregation with RDMA.** (1) Page-based system [1, 14-16]. Infiniswap [16], for example, implements a remote memory paging system based on RDMA, with a block device used as the swap space and a daemon that manages remotely accessible memory. Such systems rely on the virtual memory subsystem (VMS) and are transparent to applications, but they swap local memory pages with remote memory pages by triggering page faults, which affect performance significantly.

In addition, there is read/write data amplification problem in page-based systems. They also rely on write page faults for dirty data tracking, which also reduces system performance. (2) Object-based system [6, 7, 10, 13, 17, 18]. These systems can achieve finer memory management with custom memory objects to avoid page faults and read/write data amplification caused by page-based approaches. However, they need to use custom APIs and modify application codes. For example, AIFM [17] defines various APIs for remote memory allocation, free, prefetch, etc. It sacrifices transparency and increases migration costs. Clio [18] does not use RDMA, but still uses a similar DMA protocol with the same overhead such as DMA copies. It also requires custom APIs.

**Memory Disaggregation with CXL (or similar protocols).** These systems [19-23] avoid application code changes and reduce latency, compared to RDMA-based systems. DirectCXL [23] implements memory disaggregation based on CXL and uses the CXL switch to build a cluster. However, it can only interconnect within a rack or between adjacent racks, and the cluster size is limited. Kona [20] tries to use CC-FPGA and introduces the network to extend the distance between the host and the memory pool. However, Kona still relies on RDMA libs for remote data fetching, eviction, etc., and suffers from RDMA overheads. In addition, it is only implemented as a C library and does not use the actual hardware. To our best knowledge, there is still no research that uses CXL and can ignore its physical distance limitation. ThymesisFlow [21] achieves memory disaggregation on POWER9 architecture using OpenCAPI-enabled FPGA, while we use CXL-enabled FPGA on Intel Sapphire Rapids (SPR) architecture. However, it accesses the memory of another CPU instead a memory pool and will be limited when larger memory expansion is needed. ThymesisFlow does not fully consider latency optimization, while we add a cache on the critical path of memory access and design a congestion control algorithm to reduce network congestion.

*B. Network Latency Optimization*

Latency is one of the major problems to be solved for memory disaggregation. Access to remote memory is mainly implemented through high-speed data center networks, and many scholars have also performed latency optimization in this field. Zhuohui Duan et al. propose Gengar [42], a distributed memory pool system based on RDMA. By redesigning the RDMA communication protocol, they use a proxy mechanism to reduce the write latency of RDMA. In addition, some studies optimize latency by reducing the latency on the network access path. Xin Jin et al. propose the NetCache system [43], a system applied to key-value storage services, which reduces latency by adding caches to switches on the network access path. If there is a cache hit in the switch, there is no need to access the remote server, thus reducing latency.

*C. Congestion Control and Retransmission Mechanism*

There are some congestion control algorithms such as QCN [35], DCQCN [36], HPCC [37], etc. for high-performance networks in data centers. They all require the support of the switches, i.e. they require additional parameters or packets from the switches. For example, In-band Network Telemetry (INT) packets [44] are required in the HPCC algorithm to provide the timestamp and the number of bytes sent, as well as the queue length in the switch. The retransmission mechanism is implemented mainly through the ACK/NAK mechanism. The implementation of the retransmission mechanism on the data link layer can refer to the PCIe specification [39]. In addition, to reduce the number of retransmissions, the selective acknowledgment mechanism [40] is also used, such as IRN [45], a NIC based on RDMA.

III. DESIGN

*A. Overview*

We use CXL-enabled CC-FPGA, so the CPU can access the disaggregated memory directly via load/store semantics. We use Ethernet to transmit CXL memory access requests and ignore the physical distance limitation of CXL. We consider latency optimization and add cache on the critical path and design the network congestion control algorithm. Fig. 2 illustrates the architecture of *CXL over Ethernet* approach.

*B. CN Management*

We use CXL.mem and CXL.io protocol of CXL specifications for memory expansion and resource pooling scenarios [31, 32]. Our approach allows applications to access network-attached memory directly using virtual addresses. First, with the CXL.io protocol, the network-attached memory can be mapped to the host memory space through the IOMMU.

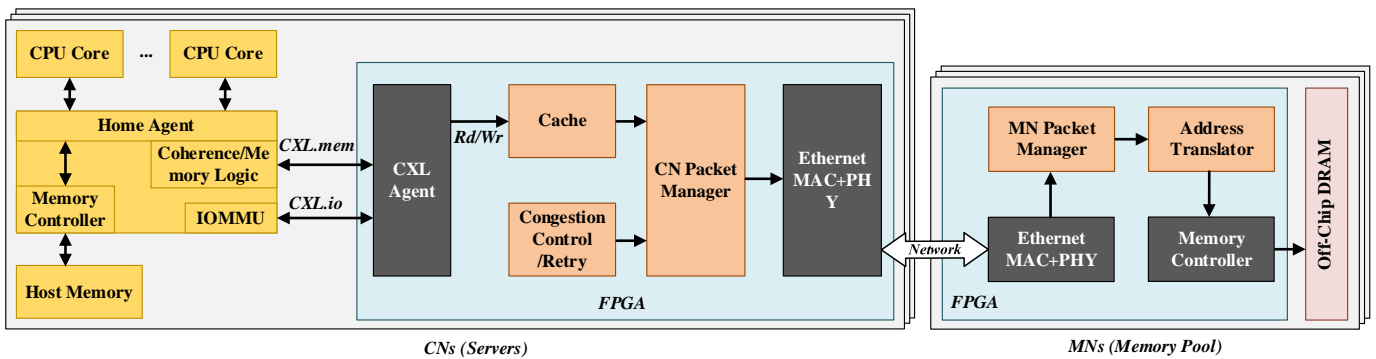

Fig. 2. The architecture of *CXL over Ethernet* approach.

For read and write instructions issued by the application, the CPU will access the physical memory using the physical address translated through TLB and MMU if a cache miss occurs. Home Agent (HA) determines if the physical address is a local memory address, and if so, accesses the host memory through Memory Controller (MC). If it is a CXL-mapped address, the CXL root port (RP) module on the CPU side encapsulates the read and write instructions as CXL flits according to the CXL-mem protocol interface (CPI) [33] and transfers them to the FPGA via the PCIe link. The CXL agent on FPGA intercepts the CXL flit and initiates the subsequent memory accesses. The access request goes through the cache, is then encapsulated to the custom formatted Ethernet frame by CN Packet Manager, and is sent to the MN via the Ethernet.

As CXL supports memory semantics, CXL memory extensions can be invoked in the same way as local memory with no software-related invocation overhead, just with increased PCIe link and Ethernet transfer latency. PCIe latency cannot be avoided, so network transmission latency is the critical part of total latency. Therefore, we add cache on FPGA to cache the remote memory data and reduce the Ethernet usage, which can effectively reduce the average latency. The cache is transparent to the server and is not shared among CNs. Each cache line uses only 3 states of the classic MESI protocol [34], i.e., modified M-state, exclusive E-state, and invalid I-state. When there is a cache hit, the CXL memory request can be immediately returned instead of using Ethernet to access remote memory. When there is a cache miss, Ethernet is usually used. However, if it is a write miss and the cache is not full, the data is simply written into the cache and the request then can be returned, unlike the typical write-back cache which usually uses the write-allocate policy. In this case, there is no need to notify the remote memory that data has been written because each CN does not share the same memory region.

### C. MN Management

After receiving access requests from CN, MN Packet Manager parses these requests, then MN can access memory devices based on the parsed results and returns data.

**Memory Allocation.** The remote memory pool is managed centrally by *Global Memory Manager (GMM)* whose functions include allocating memory and recording which CN node the memory has been allocated to. The memory is allocated at a large granularity to meet the demand. When the allocated physical memory is not enough, the CXL driver at CN will remap the address space, and GMM will reallocate the memory and update the records.

**Address Translation.** The physical memory that is mapped to the CPU address space is noted as CMem and the real physical memory of the memory pool is noted as MPMem. The Address Translator at MN is responsible for the address translation between CMem and MPMem. We implement TLB on MN FPGA (accessed by CAM), using CN id (MAC address of the CN, etc.) and CMem address for fast indexing. The complete page table is kept in the memory pool and indexed by the hash value of CN id and CMem address to improve the lookup efficiency. If the address translation is performed at CN side, CN needs to know the MPMem address, which increases the

communication between CN and MN. The OS at CN also needs to manage this process, increasing system overhead.

### D. Network Design

We offload the congestion control algorithm and the retransmission algorithm on FPGA. With the congestion control algorithm, we can reduce the latency increase caused by network congestion. The algorithm is switch-independent and can be deployed flexibly, unlike switch-dependent algorithms such as QCN [35], DCQCN [36], and HPCC [37].

**Congestion Control Algorithm.** Considering most switches are enabled with Priority-based Flow Control (PFC) [38], we use PFC packets for congestion control. The traditional PFC algorithm stops the network flow when receiving a PFC packet, so the switch queue is not fully used. Our algorithm overcomes this drawback by slowly reducing the sending rate and performing rate recovery later.

Fig. 3 shows the state transfer in our congestion control algorithm. a ~ f in Fig. 3 in turn represent the Stable Running phase, PFC Response phase, Fast Recovery phase, Fast Recovery - PFC Response phase, Increment Exploration phase, and Increment Guessing phase. The system is initially in the Stable Running phase. If a PFC packet is received, it enters the PFC Response phase, records the current rate as the target rate (TR), then decreases the rate to 1/2, and starts counting down the time (t1) to enter the next phase. If a PFC packet continues to be received, the rate is still halved and the counting starts again. If a PFC packet is received continuously within a short period (t2), it is considered a duplicate PFC packet and is ignored. It enters the Fast Recovery phase after the countdown. In this phase, if no PFC packet is received, a speedup is performed at intervals (t3), changing the rate to (CR+TR)/2, where CR is the current rate, and the speedup is considered complete after five times, then it returns to the Stable Running phase. If a PFC packet is received during the speedup, then it is considered that the rate at this point exceeds the stable rate allowed by the link, and the stable rate needs to be explored, so it enters the Fast Recovery - PFC Response phase. At this time, TR is updated to 7/8 of the current rate, while the rate is reduced to 3/4, and a countdown (t1) is started. If PFC packets are still received subsequently, continue to reduce the rate to 3/4 and restart the counting. When the counting is done, it re-enters the Fast Recovery phase, so that the stable rate can be adjusted to slightly less than the TR of the previous Fast Recovery phase. Finally, after running in the Stable Running phase for a while (t4), it enters the Increment Exploration phase to explore the maximum bandwidth allowed by the link. During the Increment Exploration phase, if no PFC packet is received, then the rate is increased by 1 Gbps at a fixed time interval (t5). Otherwise, the increase is rolled back once, then it enters the Increment Guessing phase. After a fixed time (t6), if no PFC packet is received, it is considered that the threshold value at this point is reached and it returns to the Stable Running phase. Otherwise, it is considered that the network condition becomes worse. After recording the current rate, it enters the PFC Response phase and continues to execute the algorithm. We choose suitable values for each time parameter based on simulation experiments.

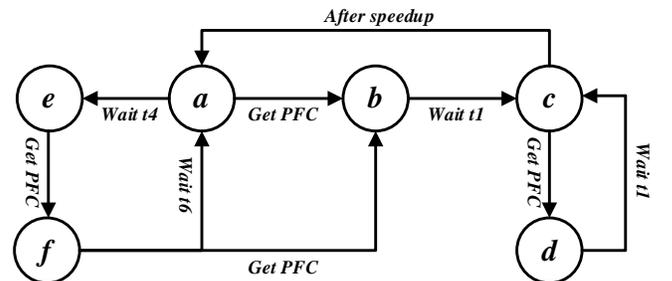

Fig. 3. The state transfer of our congestion control algorithm.

**Retransmission Mechanism.** The retransmission is based on the ACK/NAK policy [39], and each packet contains sequence numbers. We implement a retry buffer on the transmitting side of both CN and MN. There is a reorder buffer on each receiving side to keep the order of accesses for correctness. Retransmitting when the acknowledged packet times out is still a basic policy, and we also make some optimizations to reduce the number of retransmissions.

First, consider the case where MN detects a packet loss or receives a CRC-error packet. MN will respond with a NAK packet. However, to reduce the number of packets in the link to increase the valid bandwidth, we will not return this NAK immediately but wait until the subsequent request arrives and send a SACK+NAK packet. SACK indicates Selective Acknowledge [33], which is used to inform the sender that the corresponding packet has been received and can be skipped during retransmission. In addition, if there is an ACK being sent, we will also combine it and send a SACK+NAK+ACK packet. It is worth noting that since accessing memory is latency-sensitive, the delayed answer mechanism, i.e., merging multiple normal ACK packets, cannot be used. The normal ACK packet is returned immediately and is chosen to be merged with other SACK or NAK packets, as appropriate. CN then acts based on these identifiers in the packet. CN marks the packet with the corresponding sequence number in the retry buffer whenever it receives a SACK and then retransmits the packet between the marked packet and the last packet that has been acknowledged. If there is more than one SACK mark, then retransmit all packets between the current SACK mark and the previous one. The selective acknowledgment implemented in this way reduces the number of retransmissions compared to the Go-Back-N mechanism.

Second, consider the case where CN detects a packet loss or receives a CRC-error packet. Since the memory access requests are in order, the returned acknowledged packets should also be processed in order, so MN should also maintain a retry buffer for the acknowledged packets. If the subsequent ACK received at CN is correct, then immediately retransmit all request packets in which sequence numbers are between this correct ACK and the previous one, otherwise wait for the timeout to retransmit. If MN receives a request packet in which the sequence number is smaller than the currently processed sequence number, then it can be determined that there is an acknowledged packet missing or in error, so MN can immediately retransmit the corresponding acknowledged packet from the retry buffer. In this way, our approach allows fast retransmission of the acknowledged packet.

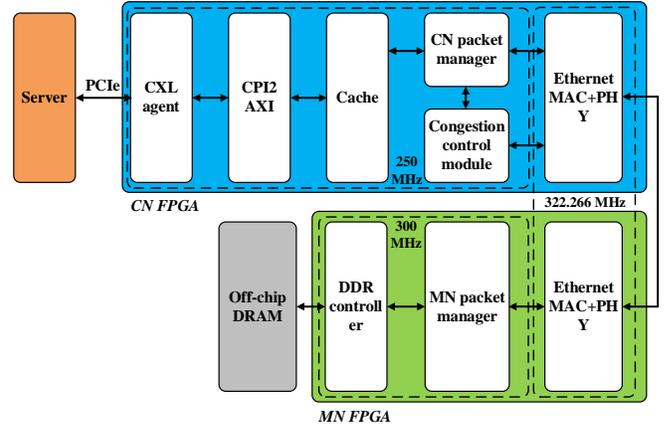

Fig. 4. FPGA implementation architecture of our approach.

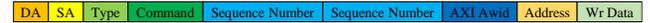

Fig. 5. Format of write request packet.

## IV. IMPLEMENTATION

Fig. 4 shows the architecture of our prototype system. We use two Xilinx U280 FPGAs, one connected to the server representing CN and the other equipped with DDR representing MN. CXL Agent on CN FPGA intercepts the CXL request from the server and converts the request to AXI bus signals. If the request is not hit in the cache, the CN packet manager encapsulates the request into an Ethernet frame and sends it to MN. MN packet manager extracts the address, and data from the packet and accesses the DDR. The result is returned to the server along the previous path. We use vendor-supplied interconnection and DDR controller IP, Ethernet MAC PHY IP, as well as CXL 1.1 standard IP. The CXL IP is implemented on FPGA as soft logic with 250 MHz clock frequency and utilizes PCIe Gen4 x8 configuration. Each module on the data path runs at 512-bit data width. We use Xilinx UltraScale+ Integrated 100G Ethernet IP running at 322.266 Mhz. MN FPGA is equipped with 32 GB DRAM and we use Xilinx DDR4 MIG IP as the DDR controller running at 300 MHz.

**Interconnection.** The CXL IP converts the CPI interface [33] into the AXI interface, which means the read/write requests to CXL Agent are then converted into AXI read/write channel signals. The DDR controller at MN is accessed through the AXI interface, so we encapsulate AXI channel signals into Ethernet frames. We customize seven packet formats, divided into request and response packets. Request packets have two formats, corresponding to read and write requests. Response packets have formats designed for congestion control and retransmission, in addition to read and write responses. Fig. 5 shows the format of a write request packet, excluding preambles and CRC, a total of 89 bytes. The Command field distinguishes packet types and the sequence numbers are used for retransmission. The Wr Data can be written to memory according to the AXI Awid and Address after MN receives the request.

**Memory.** When CXL IP performs the conversion to the AXI interface, it limits the maximum inflight number of requests, i.e. 256 read and write requests each. We determine the maximum

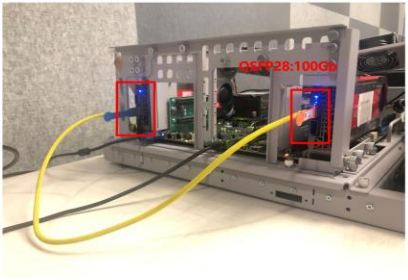
Fig. 6. Experiment environment: two boards representing MN and CN are connected via Ethernet while CN board is connected to server.

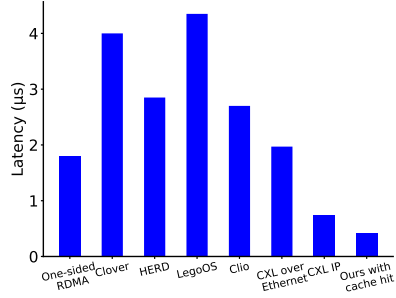
Fig. 7. Memory access latency of 64 bytes data for different approaches.

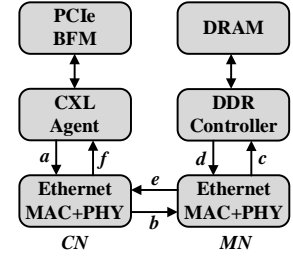
Fig. 8. Every part in access path. *a, b, c* represent request path. *d, e, f* represent response path.

size of FIFOs, retry buffers, and other memory modules based on it. For example, we set the depth of the asynchronous FIFO to 512. To reduce latency, a 4-way set associative cache with a capacity of 32 KB is integrated on CN FPGA and uses the LRU algorithm for cache line replacement. To increase the bandwidth of DRAM at MN, we use various approaches, including changing the address mapping approach to the RCB (Row-Column-Bank) approach and using address interleaving to access DRAM.

**Network.** Ethernet IP supports the PFC algorithm, so when it receives a PFC packet, our congestion control algorithm will be triggered, the execution flow is described in detail in section III. We implement a token bucket for the congestion algorithm to adjust the sending rate. The time parameters of the algorithm will be determined after experiments.

## V. EVALUATION

This section shows the onboard and simulation results of our *CXL over Ethernet* approach. Currently, we focus on round-trip memory access latency because it is the most crucial indicator of memory disaggregation. The experiment consists of a CN and an MN (Xilinx U280 boards), and the CN FPGA is connected via the PCIe Gen4 x8 link to the server equipped with an Intel SPR CPU. The OS running on the server is Fedora with kernel version 5.16.15. Fig. 6 shows our real test environment and the left board represents CN. MN and CN are interconnected via 100 Gbps QSFP28 ports on FPGA. Currently, we put CN and MN FPGA boards into the same server for equipment convenience.

**Network Path.** We first test the latency when the cache is not added on CN FPGA. We use the Intel Memory Latency Checker (MLC) [41] for latency measurements. We initiate the same number of reading and writing threads so that the read and write requests are 1/2 each. We measure the average latency of reading and writing 64 bytes, and the result is 1.97 μs. We compare it with one-sided RDMA, and some other memory disaggregation systems, including Clover [13], HERD [6], LegoOS [14], and Clio [18]. Fig. 7 shows the results. Overall, the latency of these systems is 3.14 μs on average, and the latency of our approach (*CXL over Ethernet*) is about 37% lower than it. Separately, the latency of our approach is close to one-sided RDMA and less than other disaggregation systems. Considering that the CXL IP we use is implemented as soft logic

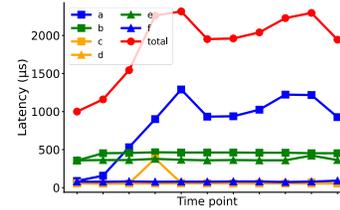
Fig. 9. Write latency of every part as well as total latency.

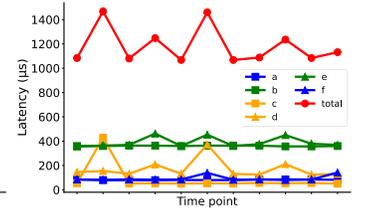
Fig. 10. Read latency of every part as well as total latency.

on FPGA, the latency can be reduced by about 300 ns if hard logic and PCIe Gen5 are used [32], then the latency of our approach will be less than one-sided RDMA.

We perform simulation tests to observe the latency of each module better. Fig. 8 shows the connection of some modules in the simulation. The notation *a~f* indicate each part of the access path. We use PCIe BFM (Bus Functional Model) to interact with the CXL agent and send access requests. Part *a* latency starts when the CXL agent sends the request and ends when CN MAC receives the sending packet. Part *b* latency starts when CN MAC receives the sending packet and ends when MN MAC receives the packet and indicates the processing time of the Ethernet IP at both sides. Part *c* latency starts when MN MAC receives the packet and ends when the DDR controller accesses DRAM. Part *d, e, f* are the opposite of part *c, b, a*.

First, we test the write latency by sending a total of 10K write requests. We evenly select 11 time points and record the latency of a request that passes through part *a~f* at each time point, as shown in Fig. 9. The simulation results show the minimum access latency is about 1 μs, while the average latency is 1.88 μs. There is a big latency difference between the first request and the subsequent requests, which is caused by the latency increase of part *a*. It is because there are fewer requests in the link at the beginning, and as requests increase, CN MAC blocks the flow due to backpressure, so requests stay in FIFO for a longer time, increasing part *a* latency. Except for part *a* latency, part *b* and *e* latency, i.e., Ethernet IP processing time, also accounts for a large part of total latency. Second, we test the read latency by sending 10K read requests, and the results are shown in Fig. 10. The minimum latency is 1.08 μs and the average latency is 1.18 μs. Unlike the write test, part *a* latency does not increase

TABLE I. MEMORY ACCESS LATENCY DIFFERENCE ON DIFFERENT FPGAS

|  | Xilinx U280 | Intel Agilex | Host DDR |
|---|---|---|---|
| Local Access Latency (ns) | 735 | 418 | 94 |
| Remote Access Latency (ns) | 850 | 615 | 198 |
| PCIe Configuration | PCIe Gen4 x8 | PCIe Gen5 x16 | None |
| CXL IP | PCIe IP and Soft Synthesized Logic | Hardened CXL IP on the R-tile | None |
| DDR Memory on Board | DDR4 with 2400 MT/s | DDR4 with 2666 MT/s | DDR5 with 4800 MT/s |

TABLE II. LATENCY RESULTS WHEN CACHE INTEGRATED

| Case | Write hit | Read hit | Write miss but not replace | Write miss and replace 1 | Read miss[1] | Write miss and replace 2 | Read miss[2] |
|---|---|---|---|---|---|---|---|
| Latency (cycle) | 14 | 16 | 14 | 268 | 285 | 36 | 50 |

1. Access to remote memory over Ethernet, 2. Access to extended memory on local FPGA.

significantly, indicating there is less CN MAC backpressure. The reason is that read requests do not carry data, so Ethernet IP processes less data within its processing ability. In addition, part *b* and *e* latency are still the major part of total latency.

In summary, the simulation results show that the latency bottleneck is network processing and transmission latency, which accounts for 70% of total latency. Therefore, we add cache to CN FPGA and solve the memory access at CN side as much as possible to reduce the network latency.

**Native CXL IP Path.** To investigate the performance difference caused by CXL IP implementation on different FPGA boards, we evaluate the local memory expansion on Xilinx U280 FPGA with soft CXL IP and Intel Agilex FPGA with hardened CXL IP respectively. We perform MLC tests on the same SPR CPU server and OS kernel respectively and the result is summarized in Table I. The local access means the FPGA memory expansion and the CPU are within the same NUMA node while the remote access means the CPU and the FPGA boards are in different NUMA Node and thus the access need to pass Ultra Path Interconnect (UPI) and then obtain the memory information, resulting in longer access path and latency. It is seen that the memory access can be further lowered by about 300 ns with PCIe Gen 5, hardened CXL IP, and higher speed DDR memory, reaching 418 ns and 615 ns for local and remote access respectively. Therefore, we believe that our approach will have lower latency than one-sided RDMA latency if implemented on Intel Agilex FPGA.

**Cache.** We perform onboard and simulation experiments to test the latency improvement after adding cache to CN FPGA. The current cache size is 32 KB, and we especially design a scenario to measure the round-trip latency with all cache hit and MLC tool. The tested average result is about 415 ns after adding cache block and is also shown in Fig. 7. Compared to the 735 ns of accessing DRAM on local FPGA through CXL and 1.97 μs of accessing remote memory via *CXL over Ethernet*, the optimized result can further reduce 300 ns and 1.5 μs, improving 41% and 76% respectively. Therefore, it can be inferred that the average access latency of our *CXL over Ethernet* approach will be improved in more scenarios by using the cache mechanism on CN FPGA.

We perform software simulation tests to see more precisely the improvement from adding cache and Table II shows the latency results. The simulation platform is the same as that shown in Fig. 8, except that the cache is added between the CXL agent and Ethernet IP, and the latency measurement starts when the CXL agent sends a request and ends when it receives a response. The cache hit latency is approximately 18.4x and 2.9x better than accessing remote memory via Ethernet and accessing local FPGA DRAM, respectively. Therefore, adding the cache can greatly reduce the average latency of *CXL over Ethernet* approach.

**Congestion Control.** We offload our switch-independent congestion control algorithm to FPGA to enable the entire design with normal switches since QCN switches have not been widely used. In general, switch-based algorithms can obtain more information from switches and are more efficient, but more difficult to deploy. We make a trade-off between efficiency and ease of use. We prove the feasibility of our algorithm by simulation and do not compare it with QCN.

We test the algorithm in different cases by generating PFC packets in the link. We set the time parameters of the algorithm, i.e., t1 = 50 μs, t2 = 10 μs, t3 = 11 μs, t4 = 200 μs, t5 = 40 μs, and t6 = 20 μs. We simulate link congestion by changing the depth threshold of the FIFO at MN receiving side. MN will send PFC packets to the link to trigger the congestion control algorithm when the FIFO usage exceeds the threshold. Fig. 11 shows the link stable rate for different FIFO thresholds with the same initial rate of 100 Gbps. The first stable rate means the rate when the link reaches its first stable state after receiving PFC packets, and the final stable rate means the rate that the link finally reaches. After reaching the first stable state, our algorithm performs a slight rate adjustment to explore the maximum rate the link allows at this time. The link takes less time to finally reach the stable rate if the first and the final stable rate are close, indicating the efficiency of the algorithm. The final stable rate varies at different FIFO thresholds. As the FIFO threshold increases, the stable rate also gradually increases, which indicates that the bottleneck limiting the link bandwidth at this time is the FIFO depth. Corresponding to the actual situation, this is the queue length of the switch. The average difference between the first and the final stable rate is 7%, indicating good efficiency after the first rate adjustment. In addition, when we set the FIFO depth threshold to 105 entries, no PFC packet will be generated, and the link bandwidth remains at 100 Gbps. It means that 105 FIFO entries are enough for the link to avoid congestion, and our system can maintain 100 Gbps throughput without link congestion.

Fig. 12 shows the final stable rate at different initial rates with the same FIFO threshold of 50 entries. The final stable rate at different initial rates is almost the same (41~42 Gbps). It is also the maximum bandwidth the link allows at this time, which indicates that our congestion control algorithm can precisely adjust the rate to reach the maximum bandwidth depending on the link congestion.

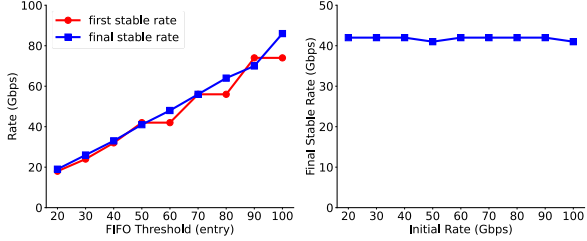

Fig. 11. Stable rates in different FIFO thresholds.

Fig. 12. Final stable rates in different initial rates.

TABLE III. RESOURCE UTILIZATION FOR U280 BOARDS AT 250 MHZ

| Resource | CXL over Ethernet with two boards | | | Native CXL IP with single FPGA |
|---|---|---|---|---|
| | Native CN | CN with cache | MN | |
| LUTs | 183581 (14%) | 233041 (18%) | 27264 (2%) | 251002 (19%) |
| FFs | 135297 (5%) | 184390 (7%) | 41498 (2%) | 172782 (7%) |
| BRAMs | 149 (7%) | 334 (16%) | 87.5 (4%) | 193 (10%) |
| IOs | 8 (1%) | 8 (1%) | 143 (23%) | 145 (23%) |
| MMCMs and PLLs | 1 (3%) | 1 (3%) | 5 (14%) | 4 (11%) |

TABLE IV. COMPARISON WITH PRIOR WORK

| | RDMA [1, 6, 7, 10, 13-18] | Native CXL [23, 32] | CC-FPGA [19, 20] | Ours |
|---|---|---|---|---|
| Memory Access Method | IO Bus | Memory Bus | Memory Bus | Memory Bus |
| CC Protocol | None | CXL: the lowest latency | CCIX: low latency | CXL: the lowest latency |
| Access Granularity | Page or custom object | Cache line | Cache line | Cache line |
| SW API Usage | Massive code refactoring with RDMA or custom APIs | Application transparent | Application transparent | Application transparent |
| SW Overhead | Rely on RDMA user space API and driver | None (page fault is exception) | Some cache operations rely on RDMA libs | None (page fault is exception) |
| Physical Link | Ethernet | PCIe | Ethernet | PCIe + Ethernet |
| Physical Limitation | None, support rack-level interconnection with TOR | Yes, within rack-level | None, support rack-level interconnection with TOR | None, support rack-level interconnection with TOR |
| Congestion Control | Yes | None | None | Yes |
| Latency (64B) | High: 2.7~4.35 μs | Low: Local Soft FPGA: ~750 ns Local Hard FPGA: ~400ns Local ASIC: ~200 ns | High: No hardware implementation | High but can be optimized. Without cache: 1.97 μs With cache hit: 415 ns |

**Resource Utilization.** Table III shows the FPGA resource utilization in our experiment. We also list the resource utilization of using CXL IP to access the extended memory on local FPGA.

## VI. CONCLUSION AND FUTURE WORK

In this paper, we propose *CXL over Ethernet* approach to implement memory disaggregation. We compare it with the prior work on memory disaggregation in terms of functionality and performance, as shown in Table IV. Our approach has been experimentally proven to combine the advantages of CXL and RDMA: it reuses the native memory semantics of CXL, allowing transparent access to disaggregated memory without code refactoring and OS call overhead and achieving low latency; based on intercepting and pulling away at the Ethernet layer, it ensures the interconnection of any switch, TOR and other network devices in the data center, greatly expanding the physical range of current CXL memory pooling and making up for the shortcomings of RDMA and native CXL approaches.

In the future, we can continue to explore memory pooling scenarios in memory disaggregation based on the current approach. In addition, we will continue to optimize the access latency with cache prefetching and hard CXL IP on FPGA. We will test more real workloads, e.g. latency-sensitive and latency-insensitive use cases, and believe our approach is competitive in many scenarios in data centers.